\documentstyle[preprint,eqsecnum,tighten,aps]{revtex}
\input epsf 

\begin{document}

\title{\hspace{1.2cm}
     Measurement of the $W$ Boson Mass }

%
%
%
\author{                                                                        
S.~Abachi,$^{14}$                                                               
B.~Abbott,$^{28}$                                                               
M.~Abolins,$^{25}$                                                              
B.S.~Acharya,$^{43}$                                                            
I.~Adam,$^{12}$                                                                 
D.L.~Adams,$^{37}$                                                              
M.~Adams,$^{17}$                                                                
S.~Ahn,$^{14}$                                                                  
H.~Aihara,$^{22}$                                                               
J.~Alitti,$^{40}$
% ...
\\                                                                              
\vskip 0.50cm                                                                   
\centerline{(D\O\ Collaboration)}                                               
\vskip 0.50cm                                                                   
}                                                                               
\address{                                                                       
\centerline{$^{1}$Universidad de los Andes, Bogot\'{a}, Colombia}               
\centerline{$^{2}$University of Arizona, Tucson, Arizona 85721}                 
\centerline{$^{3}$Boston University, Boston, Massachusetts 02215}               
\centerline{$^{4}$Brookhaven National Laboratory, Upton, New York 11973}        
\centerline{$^{5}$Brown University, Providence, Rhode Island 02912}             
\centerline{$^{6}$Universidad de Buenos Aires, Buenos Aires, Argentina}         
\centerline{$^{7}$University of California, Davis, California 95616}            
\centerline{$^{8}$University of California, Irvine, California 92717}           
\centerline{$^{9}$University of California, Riverside, California 92521}        
\centerline{$^{10}$LAFEX, Centro Brasileiro de Pesquisas F{\'\i}sicas,          
                  Rio de Janeiro, Brazil}                                       
\centerline{$^{11}$CINVESTAV, Mexico City, Mexico}                              
\centerline{$^{12}$Columbia University, New York, New York 10027}               
\centerline{$^{13}$Delhi University, Delhi, India 110007}                       
\centerline{$^{14}$Fermi National Accelerator Laboratory, Batavia,              
                   Illinois 60510}                                              
\centerline{$^{15}$Florida State University, Tallahassee, Florida 32306}        
\centerline{$^{16}$University of Hawaii, Honolulu, Hawaii 96822}                
\centerline{$^{17}$University of Illinois at Chicago, Chicago, Illinois 60607}  
\centerline{$^{18}$Indiana University, Bloomington, Indiana 47405}              
\centerline{$^{19}$Iowa State University, Ames, Iowa 50011}                     
\centerline{$^{20}$Korea University, Seoul, Korea}                              
\centerline{$^{21}$Kyungsung University, Pusan, Korea}                          
\centerline{$^{22}$Lawrence Berkeley National Laboratory and University of      
                   California, Berkeley, California 94720}                      
\centerline{$^{23}$University of Maryland, College Park, Maryland 20742}        
\centerline{$^{24}$University of Michigan, Ann Arbor, Michigan 48109}           
\centerline{$^{25}$Michigan State University, East Lansing, Michigan 48824}     
\centerline{$^{26}$Moscow State University, Moscow, Russia}                     
\centerline{$^{27}$University of Nebraska, Lincoln, Nebraska 68588}             
\centerline{$^{28}$New York University, New York, New York 10003}               
\centerline{$^{29}$Northeastern University, Boston, Massachusetts 02115}        
\centerline{$^{30}$Northern Illinois University, DeKalb, Illinois 60115}        
\centerline{$^{31}$Northwestern University, Evanston, Illinois 60208}           
\centerline{$^{32}$University of Notre Dame, Notre Dame, Indiana 46556}         
\centerline{$^{33}$University of Oklahoma, Norman, Oklahoma 73019}              
\centerline{$^{34}$University of Panjab, Chandigarh 16-00-14, India}            
\centerline{$^{35}$Institute for High Energy Physics, 142-284 Protvino, Russia} 
\centerline{$^{36}$Purdue University, West Lafayette, Indiana 47907}            
\centerline{$^{37}$Rice University, Houston, Texas 77251}                       
\centerline{$^{38}$Universidade Estadual do Rio de Janeiro, Brazil}             
\centerline{$^{39}$University of Rochester, Rochester, New York 14627}          
\centerline{$^{40}$CEA, DAPNIA/Service de Physique des Particules, CE-SACLAY,   
                   France}                                                      
\centerline{$^{41}$Seoul National University, Seoul, Korea}                     
\centerline{$^{42}$State University of New York, Stony Brook, New York 11794}   
\centerline{$^{43}$Tata Institute of Fundamental Research,                      
                   Colaba, Bombay 400005, India}                                
\centerline{$^{44}$University of Texas, Arlington, Texas 76019}                 
\centerline{$^{45}$Texas A\&M University, College Station, Texas 77843}         
}                                                                               

\date{\today}

\maketitle

\begin{abstract}
A measurement of the mass of the $W$ boson is presented based 
on a sample of 5982 $W \rightarrow e \nu$ decays observed 
in $p\overline{p}$ collisions at $\sqrt{s}$ = 1.8~TeV with the D\O\ detector
during the 1992--1993 run. 
From a fit to the transverse mass spectrum, 
combined with measurements of the $Z$ boson mass, 
the $W$ boson mass is measured to be 
$M_W =  80.350 
              \pm 0.140 ~\rm{\left (stat. \right )} 
              \pm 0.165 ~\rm{\left (syst. \right )} 
              \pm 0.160 ~\rm{\left (scale \right )} 
                  ~\rm {GeV/c^2}$.

\end{abstract}



The parameters of the gauge sector of the electroweak Standard Model~\cite{sm}
  can be taken to be the fine structure constant, the Fermi constant, 
  and   the mass of the $Z$ boson, $M_Z$, all measured to a precision 
  better than 0.01\%. 
Higher order calculations then relate the mass of the 
  $W$ boson, $M_W$, and the weak mixing
  angle, $\theta_W$, to these three parameters, the 
  heavy fermion masses, and the Higgs boson mass. 
Within the Standard Model, 
  a direct measurement of $M_W$ thus constrains
  the allowed region for the top quark and Higgs masses.
Alternatively, a precision measurement of the $W$ mass, 
  when combined with other measurements of $\sin^{2}\theta_W$, 
  provides a test of the Standard Model.
The mass of the $W$ boson has been measured recently in a number of
  experiments~\cite{mw_recent}. 
We present here a new precision measurement.

We have analyzed a sample of $W \rightarrow e\nu$ decays 
  resulting from $p\bar p$ collisions at $\sqrt{s}$ = 1.8~TeV.  
This sample, which corresponds to an exposure of $\simeq$ 12.8~pb$^{-1}$, 
  was collected with the D\O\ detector during the
  1992--1993 run at the Fermilab Tevatron collider. 
Two components of the detector~\cite{dzero} are most 
  relevant to this analysis. 
The central tracking system is used to reconstruct 
  charged particle tracks and the interaction vertex.
A central and two end uranium liquid-argon
  calorimeters measure the energy flow over a 
  pseudorapidity range $|\eta| \leq 4.2$ \cite{def_eta}.

Both $W\rightarrow e\nu$ and $Z\rightarrow e^+ e^-$ 
  decays are used in the  analysis. 
The electrons from these decays
  tend to be isolated and of high transverse momentum, $p_T$. 
At the trigger level~\cite{xs_prl}, 
  $W$ candidates were required to have an electromagnetic (EM)
  energy cluster  with 
  transverse energy $E_T = E\sin\theta \geq$ 20~GeV
   and to have missing transverse   energy 
  {\hbox{$\rlap{\kern0.20em/}E_T$}}$\geq$ 20~GeV.
Here {\hbox{$\rlap{\kern0.20em/}\vec E_T$}}$ 
     = -\sum_{\rm i} {\vec E}_{T_{\rm i}} $, 
  with the sum extending over all calorimeter cells. 
$Z$ candidates were required to have two EM energy clusters,
  each with $E_{T} \geq 10~$GeV.

Offline selection criteria were imposed 
  on the EM energy cluster of each electron candidate. 
The transverse and longitudinal shower profiles of the cluster
  were required to be consistent with those expected for an 
  electron~\cite{e_id}.
The energy leakage of the cluster into the hadronic compartment of 
  the calorimeter was required to be less than 10\%.
The isolation criterion of the cluster was satisfied by 
  requiring the total energy within a cone of radius $R = 0.4$~\cite{def_R}, 
  centered on the  electron direction, but outside the EM core of the shower
  ($R = 0.2$), to be to be less than 15\% of the energy in the EM core. 
A spatial match of the cluster with a central detector track 
  was required. 
Electrons with cluster position in the region between the  
  cryostats $(1.2 < |\eta| < 1.5)$ or within 10\% of the 
  boundary of a calorimeter module in the central region were eliminated
  from the data sample.

Having found events with well-identified, isolated electrons and
   for $W$ bosons  the required 
   {\hbox{$\rlap{\kern0.20em/}E_T$}}, kinematic constraints were 
  imposed on the data.
The $E_T$'s of each electron in $Z$ events and of the electron 
  and neutrino in $W$ events were required to exceed 25 GeV. 
The neutrino $E_T$   was equated to the {\hbox{$\rlap{\kern0.20em/}E_T$}}.
In addition, the transverse momentum of the $W$ boson, $p_T^W,$
  had to be less than 30~GeV/c. 
These selection criteria yielded 7234 
  $W\rightarrow e\nu$ events with the 
  electron in the central calorimeter ($|\eta| < 1.2$), 
  366 $Z \rightarrow ee$ events with both electrons in the 
  central calorimeter, 
  and 281 $Z \rightarrow ee$ events with one electron in the central 
  and one in an end calorimeter ($1.5 < |\eta | < 2.5$). 

Since the longitudinal component of the neutrino momentum is not measured, 
  the $W$ invariant mass cannot be reconstructed. 
Rather, the   mass of the $W$ boson is extracted from the distribution in 
  transverse mass, defined as
  $ m^{2}_{T} = 2 \, |\vec{E}_{T}^{\,e}|\, |\vec{E}_{T}^{\,\nu}|
  \, (1-\cos\varphi_{e \nu})$,
  where $\varphi_{e \nu}$ is the angle 
  between the electron and neutrino transverse momenta.
The electron direction is defined using the 
  centroid of the calorimeter cluster and the weighted average of the
  $z$ positions of the hits on the track.
The uncertainty in determining this angle leads to an uncertainty of
  $50~\rm {MeV/c^2}$ on $M_W$.
Since the absolute energy scale of the EM calorimeter
  is not known with the required precision, 
  the ratio of the measured $W$ and $Z$ masses and the world average 
  $Z$ mass~\cite{Zmass} were used to determine the $W$ boson mass. 
  The module-to-module calibration of the central EM calorimeter
  was determined to a precision of 0.5\%.
The energy resolution of the central EM calorimeter has 
  been parametrized for this analysis as 
  $\sigma/E = 0.015 \oplus 0.13/\sqrt{ E_T }\oplus 0.4/E$, 
  with $E$ in GeV. 
The sampling term of $0.13/\sqrt {E_T}$ was measured in a test beam; 
  the constant term of $0.015^{+0.006}_{-0.015}$ was determined directly from 
  the observed width of the $Z$ resonance. 
The uncertainty in the EM energy 
  resolution contributes a 70~MeV/c$^2$ uncertainty on $M_W$.

\begin{figure}[tp]
  \centerline{\epsffile{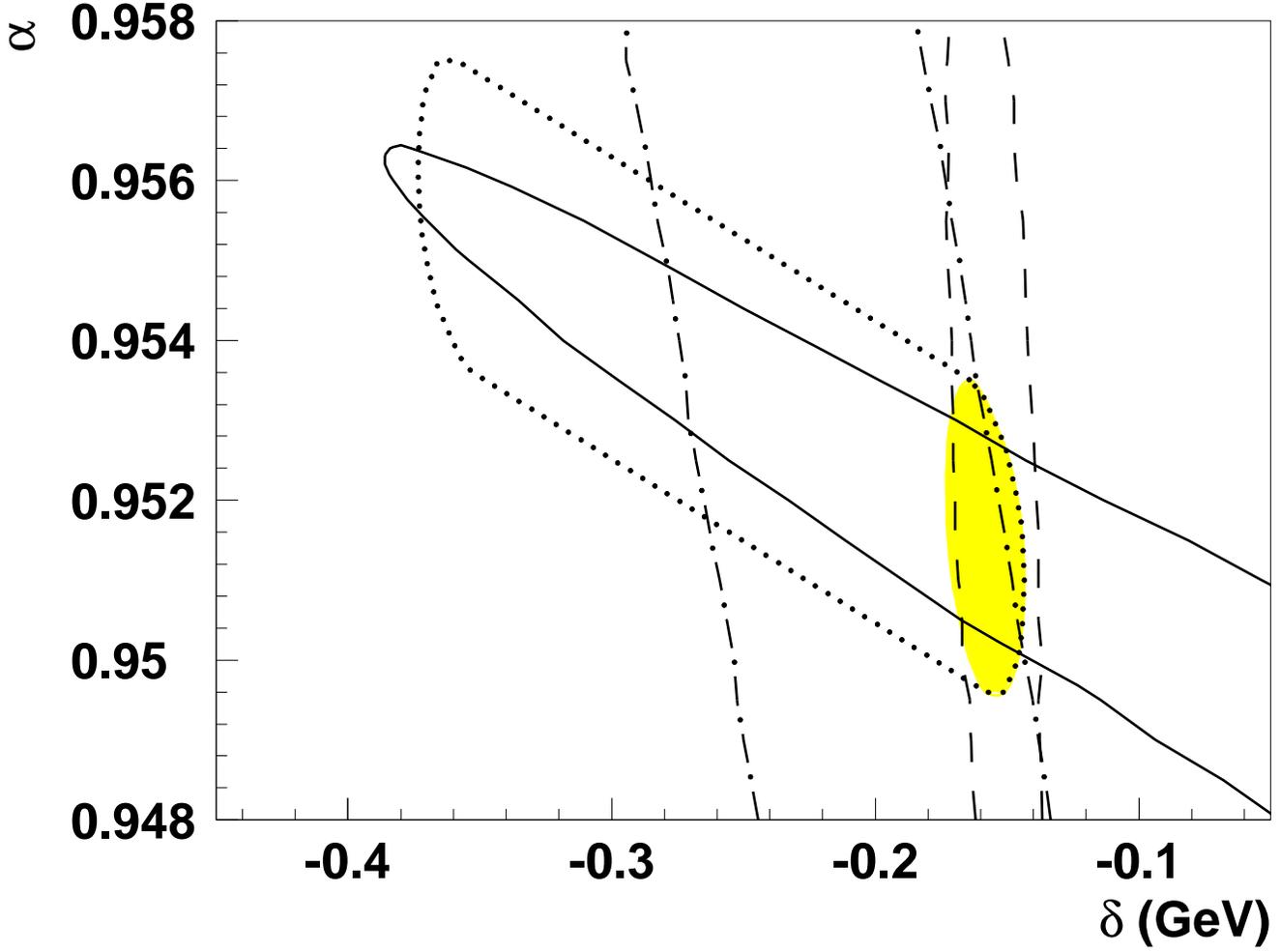}}
  \caption[]{Constraints on slope $\alpha$ and intercept $\delta$ from 
           observed $J/\psi \rightarrow e^+e^-$ (dashed-dotted line), 
           $\pi^0 \rightarrow \gamma\gamma$ (dashed line), 
           and $Z \rightarrow e^+e^-$ decays (solid line). 
           The shaded inner contour shows the combined result. 
           The dotted line indicates the allowed area when 
           nonlinear terms, as constrained by test beam measurements,
           are included. }
  \label{fig:contour}
\end{figure}

The EM energy scale of the central calorimeter was determined by
  comparing the masses measured in 
 $\pi^0 \rightarrow \gamma\gamma$, $J/\psi \rightarrow e^+e^-$, 
  and $Z\rightarrow e^+e^-$ decays to their known 
  values~\cite{Zmass,ref_mass}. 
If the electron energy measured in the calorimeter and the true energy are 
  related by
  $E_{\rm meas} = \alpha \, E_{\rm true} + \delta$, 
  the measured and true mass values are, to first order, related 
  by $m_{\rm meas} = \alpha \, m_{\rm true} \,+\, \delta \, f $. 
The variable $f$ depends on the decay topology and is given by 
  $f = {2(E_1 + E_2) \over m_{\rm meas} } \sin^2\gamma/2$, 
  where $\gamma$ is the opening angle between the two decay products 
  and $E_1$ and $E_2$ are their measured energies. 
Figure~\ref{fig:contour} shows the constraints on the parameters $\alpha$
  and $\delta$ obtained independently from the $\pi^0$, the $J/\psi$,
  and the $Z$ data. 
When combined, these three constraints 
  limit $\alpha$ and $\delta$ to the shaded elliptical region. 
Test beam measurements allow for a small nonlinear term in the 
  energy response, which affects both $\alpha$ and $\delta$
  and  alters the ratio $M_W / M_Z$ largely through the effect on $\delta,$
  as shown by the dotted line in Fig.~\ref{fig:contour}.

Using the measured masses for the observed resonances, the energy 
  scale factor determined is $\alpha =
  0.9514 \pm 0.0018 {}^{+0.0061}_{-0.0017}$ and the offset is $\delta =
  -0.158 \pm 0.015 {}^{+0.03}_{-0.21}$~GeV, where the asymmetric errors
  are due to possible nonlinearities. 
The measured offset is consistent with that determined from test beam data, 
  and has been confirmed by a detailed Monte Carlo study of 
  energy loss in the central detectors. 
The dependence of the measured ratio of the $W$ mass to $Z$ mass on 
  $\alpha$ and $\delta$ may be estimated from 
  \begin{eqnarray*}
    \left. \frac{M_W (\alpha,\delta)}{M_Z (\alpha,\delta)}\right|_{\rm meas} 
           &=&  
    \left. \frac{M_W}{M_Z}\right|_{\rm true} 
           \left[ 1 + \frac{\delta}{\alpha} \cdot 
                      \frac{f_W \, M_Z - f_Z \, M_W}{M_Z \cdot M_W} \right]. \
  \end{eqnarray*}
It should be noted that the $W$ mass is insensitive to $\alpha$ if 
  $\delta=0$. 
The uncertainty on the absolute energy scale 
  results in an uncertainty on $M_W$ of 160~MeV/c$^2$, of which 
  150~MeV/c$^2$ is due to the statistics of the $Z$ data sample.

The $W$ mass is obtained from an unbinned maximum-likelihood fit of the 
  data to distributions in $m_T$,
  generated as a function of $M_W$
  at 100~MeV/c$^2$ intervals by a fast Monte Carlo simulation. 
This Monte Carlo models both the production and decay of the 
  vector bosons
  and the detector response, and relies heavily on experimental data for input. 
It starts with the double differential $W$ production cross section in 
  $p_T$ and rapidity 
  calculated at next to leading order~\cite{ly}
  using the MRSA parton distribution functions (pdf)~\cite{mrsa}. 
The mass of the $W$ boson is generated with a 
  relativistic Breit-Wigner line shape, 
  skewed by the mass dependence of the parton luminosity.
In the simulation, the $W$ boson width 
  has been fixed to its measured value, 
   $\Gamma_W = 2.07 \pm 0.06 ~\rm{GeV/c^2}$~\cite{xs_prl}. 
The uncertainty on $\Gamma_W$ results in an uncertainty 
  of 20~MeV/c$^2$ on $M_W$. 
The $W$ decay products are then generated in the $W$ rest frame 
  with an angular distribution respecting the 
  polarization of the $W$. 
Radiative decays are generated at ${\cal O}(\alpha)$ 
according to~\cite{Berends}.

\begin{figure}[tp]
    \centerline{\epsffile{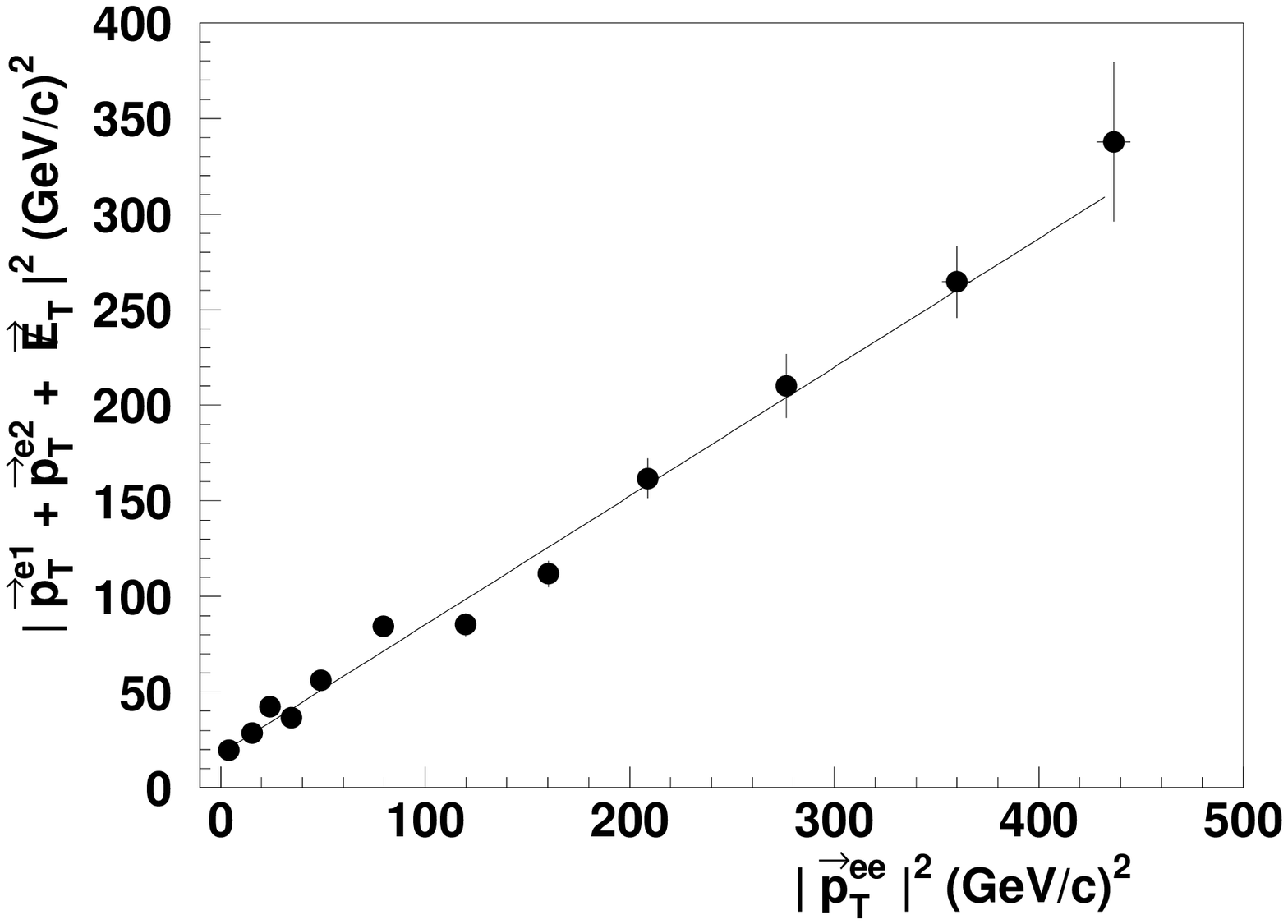}}
\caption[]{ Distribution of 
              $ | {\vec p}_T^{\,e_1} + {\vec p}_T^{\,e_2} + 
              {\hbox{$\rlap{\kern0.20em/}\vec E_T$}} |^2 $ 
              versus $ | {\vec p}_T^{\,ee} |^2 $ for $Z$ events. }
\label{fig:urecoil}
\end{figure}

After generation of the kinematics of the event at the four-vector level, 
  the resolutions of the detector are incorporated and
  the energy scales are set.
Minimum bias (MB) events are used to model the underlying event, 
  mimicking the debris in the event due to spectator parton 
  interactions and the pile-up associated with multiple interactions, 
  and including the residual energy from previous beam crossings.
The relative response of the hadronic and EM calorimeters
  is established by studying $Z$ events. 
To ensure an equivalent event topology between the $W$ and $Z$ events, 
  $Z$ decays in which one electron is in the end calorimeter are included
  in this study.
The transverse momentum balance in $Z$ events is given by  
  $ {\vec p}_T^{\,e_1} + {\vec p}_T^{\,e_2} + {\vec p}_T^{\,rec} 
                     + {\vec u_T}     
     = - {\hbox{$\rlap{\kern0.20em/}\vec E_T$}} $, 
  where ${\vec u_T}$ is the underlying event contribution and 
  ${\vec p}_T^{\,rec}$ is the transverse momentum of the recoil to the 
   vector boson.
One finds for the average
  $    | {\vec p}_T^{\,e_1} + {\vec p}_T^{\,e_2} + 
   {\hbox{$\rlap{\kern0.20em/}\vec E_T$}} |^2 = 
     \kappa^2 \, | {\vec p}_T^ {\,ee} |^2 + | {\vec u_T} |^2 $ 
  assuming
  $ | {\vec p}_{T}^{\,rec} | = \kappa \, | {\vec p}_{T}^{\,ee} | $, 
  where $ {\vec p}_{T}^{\,ee} $ is the transverse momentum of the $Z$
   measured from the two electrons. 
The cross term on the right hand side averaged to zero since 
  the underlying event vector is randomly distributed with respect 
  to the $Z$ recoil system. 
Figure~\ref{fig:urecoil} shows the distribution of 
  $ | {\vec p}_T^{\,e_1} + {\vec p}_T^{\,e_2} + 
  {\hbox{$\rlap{\kern0.20em/}\vec E_T$}} |^2 $ 
  versus $ | {\vec p}_T^{\,ee} |^2 $. 
The data shows a linear relation between the EM 
  and hadronic energy scale, and yields 
  $\kappa = 0.83 \pm 0.04$. 
The intercept yields the magnitude of the 
  underlying event vector, $|{\vec u_T}| = 4.3 \pm 0.3 ~\rm{GeV/c}$, 
  consistent with the value obtained from MB events.
The uncertainty on $M_W$ due to the uncertainty on the hadronic energy 
  scale is 50~MeV/c$^2$. 

The recoil against the vector boson is modeled by a single jet. 
The  transverse momentum of the $W$ is scaled by $\kappa$ and smeared using a 
  resolution of  0.80/${\sqrt {p_T^W \rm (GeV)} }$, as obtained 
  from our dijet events. 
The uncertainty on the jet resolution gives a 65~MeV/c$^2$ 
  uncertainty on $M_W$. 
The event is superimposed onto MB events, which simulates
  the underlying event.
The luminosity profile of these MB events is chosen such that the
  mean number of interactions per crossing is the same as for the
  $W$ data.

The modeling of the recoil and  underlying event 
  are verified and constrained 
  by comparing the $p_T$ of the $Z$ obtained from
  the two electrons, $\vec{p}_T^{\,ee}$, to that
  obtained from the rest of the event:  $-\vec{p}_T^{\,rec} - {\vec u_T}$. 
To minimize the contribution from the electron energy resolution,
  the vector sum of these two quantities is projected
  along the bisector of the two electron directions.
Since $\vec u_T$ is randomly oriented and has a magnitude 
  $\sim p_T^Z$, the width of the distribution is sensitive 
  to the underlying event contribution while the mean is largely 
  unaffected.  
The $rms$ of the distribution  is $4.44 \pm 0.18~\rm {GeV/c}$.
The sensitivity of the width of this distribution to the mean number
  of MB events that mimic the underlying event is determined
  by varying the number of MB events in the Monte Carlo.
The number of MB events preferred by the data is 
  $0.98 \pm 0.06$, consistent with one. 
The uncertainty on $M_W$ from the underlying event model is 60~MeV/c$^2$. 

The energy underlying the electron was obtained from $W$ events
  by measuring the energy deposited in a region of the calorimeter the same 
  size as the electron cluster but rotated away from the electron in azimuth. 
On average, the underlying event adds $205\pm 55 ~\rm{MeV}$
  to the energy of central electrons and results in an
  uncertainty on $M_W$ of  35~MeV/c$^2$.

Detector and reconstruction biases were also modeled in the Monte 
  Carlo simulation. 
In radiative decays, $W \rightarrow e\nu\gamma$, the $e\nu$ mass does not 
  reconstruct to the $W$ mass unless the photon is clustered with the 
  electron. 
Also, radiative decays in which the photon 
  is radiated near, but not fully within, the electron cluster 
  can distort the cluster shape causing 
  the electron to fail the shower shape cuts. 
The same considerations apply to radiative $Z$ decays and these 
  effects do not cancel completely   in the ratio of the masses. 
Similarly, the recoil system may affect the electron 
  identification, especially if it is close to the electron. 
A measure of the event selection biases, due to electron shape and isolation 
  cuts, is obtained by studying the projection 
  of the momentum recoiling against the $W$
  along the electron $p_{T}$ direction: 
  $u_\parallel \equiv (\vec{p}_T^{\,rec} + \vec u_T )\cdot \hat{p}_T^{\,e}$. 
An inefficiency in $u_\parallel$ would cause a kinematic bias for the 
  $W$ decay products. 
The efficiency as a function of $u_\parallel$ has been determined from 
  the $W$ data using the energy in a cone around the electron, which is
  used to select isolated electrons.
The efficiency was verified using $Z$ decays. 
For $u_\parallel$ values of $20~\rm {GeV}$ there is an inefficiency 
  of approximately 10\%. 
The error on $M_W$
 resulting from the uncertainty in the $u_\parallel$ efficiency
 is $20~\rm {MeV/c^2}$.

The QCD jet background in the $W$ sample was determined from an independent
  jet data sample to be $(1.6 \pm 0.8) \%$. 
Inclusion of this background shifts the mass by +33~MeV/c$^2$. 
The background from $Z\rightarrow e^+e^-$ events in which one electron is 
  not identified has been estimated, using 
  {\footnotesize ISAJET}~\cite{isajet},   to be $(0.43 \pm 0.05)$\%. 
Its effect on $M_W$ is negligible. 
The uncertainty in the amount of background, and its distribution in 
  transverse mass, gives an uncertainty on $M_W$ of 35~MeV/c$^2$. 
The 1.3\% irreducible background due to 
  $W\rightarrow \tau\nu \rightarrow e\nu\nu\nu$ was included in the Monte 
  Carlo simulation. 
All other sources of background are negligible. 

\begin{figure}[tp]
    \centerline{\epsffile{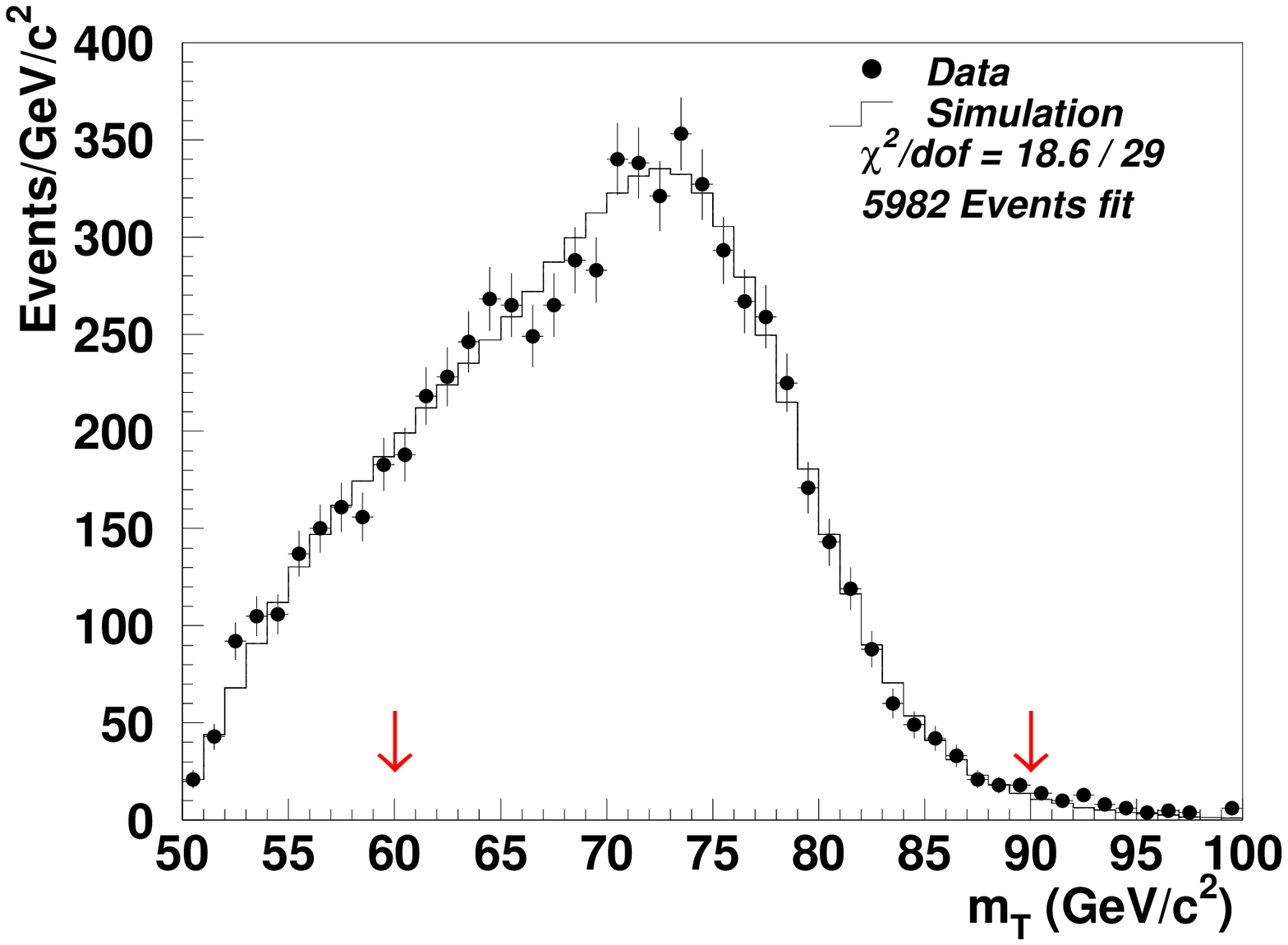}} 
\caption[]{Best fit to the transverse mass distribution. The arrows indicate 
           the fitting range from which the $W$ mass is extracted.  }
\label{fig:mt_fit}
\end{figure}

The distribution in $m_T$ and the Monte Carlo line shape 
  corresponding to the best fit are shown in Fig.~\ref{fig:mt_fit}. 
The mass, extracted from a fit of the 5982 events in the range
  $60 \leq m_T \leq 90$~GeV/c$^2$, is 
  $M_W =   80.350 \pm 0.140 \ {\rm (stat.)} \pm 0.165 \ {\rm (syst.)} 
              \pm 0.160 \ {\rm (scale)}$~GeV/c$^2$.
Table~\ref{table:sys} lists the uncertainties in the measurement, which
  used the MRSA pdf.
As a consistency check, a fit to the $p_T^e$ distribution in the range
  $30 \leq p_T^e \leq 45$~GeV/c$^2$
  was performed to extract the $W$ mass.
This fit results in a mass $50~\rm{MeV/c^2}$ lower than when measured from
  the $m_T$ distribution.
The statistical error on this fit is $190~\rm{MeV/c^2}$.

The largest systematic uncertainty, beyond those mentioned above, 
  is due to the modeling of the $p_T^W$ spectrum and the pdf's. 
The correlation between the pdf's and the $p_T^W$ distribution 
  has been addressed. 
To study the uncertainty, parametrizations of the CTEQ3M pdf were 
  obtained~\cite{cteq3m} incorporating all available data and
  with the $W$ charge asymmetry~\cite{cdf_asym}
  data points moved coherently by  $\pm$ one standard deviation,
  resulting in a  maximum allowed range of pdf's.
The parameters governing the nonperturbative
  part of the $p_T^W$ spectrum~\cite{ly_g2} were varied simultaneously, 
  as constrained by our measured $p_T^Z$ spectrum. 
The resulting variation in the spectrum leads to an uncertainty of 
  65~MeV/c$^2$ on $M_W$.

In conclusion, a new  measurement of the $W$ mass from a fit to the 
  transverse mass spectrum of $W\rightarrow e\nu$ decays has been presented. 
The $W$ mass is measured to be
    $M_W = 80.350 \pm 0.270 $~GeV/c$^2$, where all errors have been
    added in quadrature.

%
We thank the staffs at Fermilab and the collaborating institutions for their
contributions to the success of this work, and acknowledge support from the 
Department of Energy and National Science Foundation (U.S.A.),  
Commissariat  \` a L'Energie Atomique (France), 
Ministries for Atomic Energy and Science and Technology Policy (Russia),
CNPq (Brazil),
Departments of Atomic Energy and Science and Education (India),
Colciencias (Colombia),
CONACyT (Mexico),
Ministry of Education and KOSEF (Korea),
CONICET and UBACyT (Argentina),
and the A.P. Sloan Foundation.
%

\begin{table}[hp]
\begin{center}
\begin{tabular}{l|r} 
Uncertainty                                     & MeV/c$^{2}$  \\ \hline\hline
Statistical                                     & 140   \\ 
                                                &       \\ 
Energy scale                                    & 160   \\ \hline 
\qquad   Statistical                            & 150   \hspace*{1.0cm} \\
\qquad   $Z$ systematics                        &  35   \hspace*{1.0cm} \\
\qquad   Calorimeter low energy nonlinearities  &  25   \hspace*{1.0cm} \\
                                                &       \\
Other systematics                               & 165   \\  \hline
\qquad Electron energy resolution               & 70    \hspace*{1.0cm} \\ 
\qquad Jet energy resolution                    & 65    \hspace*{1.0cm} \\ 
\qquad pdf's, $p_T^W$ spectrum                  & 65    \hspace*{1.0cm} \\ 
\qquad Underlying event model                   & 60    \hspace*{1.0cm} \\ 
\qquad Relative hadronic and EM energy scale    & 50    \hspace*{1.0cm} \\ 
\qquad Electron angle calibration               & 50    \hspace*{1.0cm} \\ 
\qquad Energy underlying electron               & 35    \hspace*{1.0cm} \\ 
\qquad Backgrounds                              & 35    \hspace*{1.0cm} \\ 
\qquad Radiative decays                         & 20    \hspace*{1.0cm} \\ 
\qquad $u_\parallel$ efficiency                 & 20    \hspace*{1.0cm} \\ 
\qquad Trigger efficiency                       & 20    \hspace*{1.0cm} \\ 
\qquad $W$ width                                & 20    \hspace*{1.0cm} \\ 
\qquad Fitting error                            &  5    \hspace*{1.0cm} \\ 
                                                &       \\
Total                                           & 270   \\      
\end{tabular}
\end{center}
\caption{Uncertainties in the $W$ boson mass measurement.}
\label{table:sys}
\end{table}

\begin{references}

%
\bibitem[*]{beijing}
Visitor from IHEP, Beijing, China.

\bibitem[\dag]{ecuador}
Visitor from Univ. San Francisco de Quito, Ecuador.

\vskip 0.25cm

\bibitem{sm}
            S. Weinberg,\ Phys. Rev. Lett. {\bf 19}, 1264 (1967);
            S.L. Glashow, Nucl. Phys. {\bf 22}, 579 (1968);
            A. Salam, in {\em Elementary Particle Theory},
            ed. by N. Svartholm (Almquist and Wiksell, Sweden, 1968), p. 367;
            S.L. Glashow, J. Illiopoulos and L. Maiani, Phys. Rev. D {\bf 2}, 
            1285 (1970);
            M. Kobayashi and M. Maskawa, Prog. Theor. Phys. 
            {\bf 49}, 652 (1973). 

\bibitem{mw_recent}
            J.~Alitti {\it et al.} (UA2 Collaboration), 
            Phys. Lett. {\bf B276}, 354 (1992);
            F.~Abe {\it et al.} (CDF Collaboration), 
            Phys. Rev. Lett. {\bf 65}, 2243 (1990), 
            Phys. Rev. D {\bf 43}, 2070 (1991); 
            F.~Abe {\it et al.} (CDF Collaboration), 
            Phys. Rev. Lett. {\bf 75}, 11 (1995), 
            F.~Abe {\it et al.} Phys. Rev. {\bf D52}, 4784 (1995). 

\bibitem{dzero}
            S.~Abachi {\it et al.} (D\O\ Collaboration), 
            {\it Nucl. Instr. and Methods} {\bf A338}, 185 (1994). 

\bibitem{def_eta}
            Pseudorapidity is defined as $\eta = -\ln\tan{\theta\over 2}$  
            where $\theta$ is the polar angle with respect to the proton beam. 

\bibitem{xs_prl}
            S.~Abachi {\it et al.} (D\O\ Collaboration), 
            Phys. Rev. Lett. {\bf 75}, 1456 (1995). 

\bibitem{e_id}
            For more details, see 
            S.~Abachi {\it et al.} (D\O\ Collaboration), 
            Phys. Rev. D {\bf 52}, 4877 (1995).

\bibitem{def_R}
            $R$ is defined as $R = \sqrt {\Delta\eta^2 + \Delta\varphi^2} $
            where $\Delta \eta$ and $\Delta \varphi$ are calculated from the 
            center of the calorimeter cells with respect to the 
            ($\eta,\varphi$) position of the electromagnetic shower. 

\bibitem{Zmass}      
            We used $M_Z^{\rm LEP}$ = 91.1884 $\pm$ 0.0022~GeV/c$^2$, from 
            P.~Renton, \lq\lq Precision Tests of Electroweak
            Theories,\rq\rq\ Lepton-Photon Conference, Beijing, 
            P.R. China (1995),  OUNP-95-20.

\bibitem{ref_mass}      
            The reference mass values used are 
            $M_{J/\psi}$ = 3.09688 $\pm$ 0.00004~GeV/c$^2$ and 
            $M_{\pi^0}$ = 0.1350 $\pm$ 0.0006~GeV/c$^2$, 
            Particle Data Group, L.~Montanet {\it et al.}, 
            Phys. Rev. D {\bf 50}, 1173 (1994).

\bibitem{ly}   
            G.~Ladinsky and C.-P.~Yuan, Phys. Rev. D {\bf 50}, 4239 (1994).

\bibitem{mrsa}     
            A.D.~Martin, R.G.~Roberts and W.J.~Stirling,     
            Phys. Rev. D {\bf 50}, 6734  (1994) and
            Phys. Rev. D {\bf 51}, 4756 (1995). 

\bibitem{Berends}  
            F.~A.~Berends and R.~Kleiss,
            Z. Phys. {\bf C27}, 365 (1985).


\bibitem{isajet}
            F. Paige and S. Protopopescu,
            BNL Report no. BNL38034 (1986, unpublished), release 6.49.
            
\bibitem{cteq3m}
            H.L. Lai, {\it et al.}, Phys. Rev. D {\bf 51}, 4763 (1995).

\bibitem{cdf_asym} 
            F.~Abe {\it et al.}, (CDF Collaboration),
            Phys. Rev. Lett. {\bf 74}, 850 (1995). 

\bibitem{ly_g2} 
            The parametrization of the $p_T^W$ spectrum is most sensitive 
            to variations of the parameter $g_2$ in the nonperturbative 
            function (see~\cite{ly}). 
            The range for $g_2$ as limited by our $Z$ data 
            was $\tilde g_2-2\sigma < g_2 < \tilde g_2+4\sigma$ where 
                $\tilde g_2=0.58^{+0.10}_{-0.20}~\rm{ (GeV/\hbar c)^2}$.

\end{references}
\end{document}